\documentclass[useAMS,usenatbib]{mn2e}
\usepackage{amsmath} 
\usepackage{subfigure}
\usepackage{times}

\usepackage{graphicx}
\usepackage{times}%
\usepackage{natbib}
\usepackage{times}%
\def\sw{{\it{Swift}}}

\title[Redshift distribution and luminosity function of LGRBs]
      {Redshift distribution and luminosity function of long gamma-ray bursts from cosmological simulations}
\author[Campisi M.A. et al.]
       {M.A. Campisi$^1$\thanks{E-mail: campisi@mpa-garching.mpg.de}, 
	L.-X. Li$^2$ and P. Jakobsson$^3$\\
        $^1$ Max-Planck-Institut f\"ur Astrophysik, 
        Karl--Schwarzschild--Str. 1, D-85748 Garching, Germany \\ 
        $^2$ Kavli Institute for Astronomy and Astrophysics, Peking University, Beijing 100871, China\\
	$^3$ Centre for Astrophysics and Cosmology, Science Institute, University of Iceland, Dunhagi 5, IS-107 Reykjavik, Iceland\\
}

\begin{document}

\date{Accepted 2010 May 18. Received 2010 April 26; in original form 2009 August 21 }
\pagerange{\pageref{firstpage}--\pageref{lastpage}} 
\pubyear{2010}

\maketitle

\label{firstpage}
\begin{abstract}
We study the luminosity function (LF), the comoving rate and the detection rate of Long Gamma-Ray Burst (LGRBs) to high redshift,
using galaxy catalogues constructed by combining high-resolution N-body
simulations with semi-analytic models of galaxy formation. We assume the collapsar model and different metallicity thresholds, 
and conclude that LGRBs are not good tracers of the star formation history in the universe.
Then using the $\log N-\log P$ diagram for BATSE bursts, we determine the LF (with and without evolution with redshift) and the formation rate of LGRBs, obtaining constraints on the slope of the power-law. 
We check the resulting redshift distribution with \sw\ data updated to 2009 August, finding that models where LGRBs have as progenitors stars with $Z<0.3Z_{\odot}$ and without evolution of the LF are in agreement with the data. We also predict that there are about $\sim1\%$ of GRBs at redshift $z>6$.
\end{abstract}

\section{Introduction} 
\label{sec:intro}
Gamma-ray bursts (GRBs) are the most luminous
explosion in the Universe \citep{zhang04}. They offer exciting possibilities for studying
astrophysics in extreme conditions, e.g., radiative processes in highly
relativistic ejecta \citep[and references therein]{huang00,fan08}. Because of
their very large luminosity, GRBs represent cosmological events, which have
been detected up to $z\sim8.2$ \citep{tan09,sal09b}.

The observed distribution of the duration of GRBs is 
bimodal \citep{kou93}: long GRBs (hereafter
LGRBs) and short GRBs, depending on whether their durations are longer or
shorter than 2 seconds.
The current favorite hypotheses for their origin are that short GRBs are produced by the
merger of compact objects
\citep{li98,osh08}, while LGRBs originate from the
death of massive stars (with low metallicity), such as Wolf-Rayet stars
({\sl the collapsar
model}) \citep{Yoon_Langer_Norman_2006,yoon08, woo06b}.
Throughout this paper we will deal only with the LGRBs.
Observational data are consistent with the hypothesis of the LGRB-supernova
connection: at least some LGRBs are associated with core-collapse supernovae \citep[and references therein]{gal98,hjo03,pia06,li06,woo06b}. In addition,
all supernovae associated with GRBs are Type Ic, which supports the
hypothesis of Wolf-Rayet stars as progenitors of LGRBs. Because of their
connection with supernovae, LGRBs are potential tracers of the cosmic
star formation history (SFH) \citep{bro02,fyn06,pri06,pro07,Savaglio_2006,tot06,li08a}, 
nonetheless this connection could be non-trivial (see e.g. \cite{koc09}).

The detection of LGRBs at high redshift and the connection with Type Ic SNe, make them promising for probing the Universe. They are probably the only objects that allow us to study the cosmos {at high redshift} and the early evolution of {PopIII stars}. 
Therefore, understanding where {GRBs} are distributed in the universe and how they are connected with the star formation rate (SFR) is very {important}.
Since the launch of the \sw\ satellite \citep{geh04}, the number of GRBs with measured redshift has been greatly increased. Nonetheless to date there are only $\sim140$ \sw\ GRBs with known redshifts. This sample is still too {small} to 
constrain their luminosity function (hereafter LF).

Previous studies \citep[and others]{Porciani_Madau_2001, gue05, nat05, dai06, sal07} 
{touched} the problem of determining the redshift distribution of GRBs differently. Their method is based on constraining the GRBs distribution by assuming an average energy spectrum for all the bursts and that {GRBs} trace the evolving SFR, either with a constant or evolving LF. The redshift distribution, together with the LF, can provide important insights not only into the physics of the individual objects themselves, but also into the evolution of matter in the Universe.

One of the first pioneering {works} was Porciani $\& $ Madau (2001). 
The rate of GRB is {fitted to} observational data, using the assumption that SFR is proportional to the GRB rate \citep{hop06}. Finally {the rate is} convolved with a selection function dependent on the instrument used.
They find that the rate of {bursts} is of about 1-2 GRBs for every one million Type II SNe.
Notable recent attempts include: Guetta, Piran $\& $ Waxman (2005) explore a variety of different star-formation rate histories and GRB luminosity {functions}; Natarajan et al. (2005) additionally incorporate a simple prescription for a low-metallicity preference {of GRBs}.
Daigne et al. 2006 used Monte Carlo simulations to predict {the GRB evolution}, assuming that GRBs follow the SFR, the LF is a power-law ({independent of} redshift) and the peak energy is determinated by {two relations}. They find that:
the slope of the LF is between $1.5-1.7$, the Amati relation should be {an intrinsic relation} and,
the GRB rate density at $z=7$ is about 6-7 times larger than at $z=2$. They {also deduced} that the properties of GRBs and GRB-progenitors are redshift dependent, since the redshift distribution of \sw\ burst strongly favors their SFR3 model (see Daigne et al 2006), {although that is} an unrealistic model.\\
Salvaterra follows this approach in two different works (Salvaterra $\&$ Chincarini 2007, Salvaterra et al. 2008), assuming that GRB luminosity evolves with redshift and that GRBs form preferentially in low-metallicity environments. They use this constraint to set a robust upper limit on the {slope of} bright-end of GRB LF, finding that the number of bright {GRBs} detected by \sw\ implies that {it} cannot be very steep ($\delta<2.6$ for progenitors with $Z<0.3Z_{\odot}$). Moreover they found that assuming a threshold of $F>0.4$ [ph/cm$^2$/s], at least $\sim 5-10\%$ of all detected GRB should lie at redshift $z>5$.\\

In this work, 
we do not use a GRB comoving rate proportional to the star formation in the Universe, but only we assume that {the global rate of GRBs} per SNe is on average (over all cosmic times) of about 1 GRB event every 1000 SNe \citep{lan06}. We derive the LF and formation rate of GRBs using a catalogue of {galaxies} constructed by combining high-resolution
N-body simulations with a semi-analytic model of galaxy formation (Wang et al. 2008). We fit the observed logN-logP relation \citep{kom00} derived from {the GRB data of} the Burst And Transient Source Experiment on board the {\it CGRO} satellite (BATSE, Fishman et al. 1989) in order to constrain the free parameters of the LF. We adopt this method for three {GRB progenitor subsamples} with different cuts in metallicity, following the collapsar model, and assuming a constant and evolving LF.\\
By comparing the cumulative distribution of peak photon fluxes, \citet{dai09} recently proved that the \sw\ and BATSE samples track the same parent population of bursts. 
For this reason, following Porciani $\& $ Madau, we rely on the {GRBs} observed by BATSE as {the two} samples should have comparable LF.

The paper is organized as follows. We
present in section 2 {the simulated GRB sample} used in this work. In
section 3, we describe the method to reproduce the redshift distribution of LGRB. We
describe our results in section 4. We give our conclusions in section 5.

\section{Simulated LGRB rate }
\label{sec:sam}
In this study, we use the galaxy catalog constructed by 
\citet{Wang_etal_2008} for { simulations} with cosmological parameters from 
the third-year WMAP results. The same catalog was used in \cite{campisi09}, but we refer to \citet{Wang_etal_2008} for a discussion of the model. 
The simulation corresponds to a box
of $125\,h^{-1} {\rm Mpc}$ comoving length and a particle mass of $7.8\times 10^8\,{\rm M}_{\odot}$. The softening
length is 5 $h^{-1}$ kpc. 
Simulation data were stored in 64 outputs, {which} are
approximately logarithmically spaced in time between $z=20$ and $z=1$, and
linearly spaced in time for $z<1$. Each simulation output was analyzed with
the post-processing software originally developed for the Millennium
Simulation \citep{Springel_etal_2005}.

In order to extract from the catalog the rate of GRB events, we adopt the collapsar model
for LGRBs: all young stars with mass $>20\,M_{\odot}$ \citep{lar07} ending their life with a
supernova should be able to create a BH remnant\footnote{We also test the case with $M>30\,M_{\odot}$. We obtain that the rate of GRB in every box is very close to the rate whit $M>20 M_{\odot}$, since using a Salpeter IMF the difference between the two cases is only $0.939\times10^{-3}$.}. If the collapsar has high
angular momentum, the formation of the BH can be accompanied by a GRB
event \citep{Yoon_Langer_Norman_2006,yoon08}. As mentioned in
Sec.~\ref{sec:intro}, recent studies on the final evolutionary stages of
massive stars suggested that a Wolf-Rayet (WR) star can produce a LGRB if its
mass loss rate is small. This is possible only if the metallicity of the star
is very low. When metallicities are lower than $\sim 0.1-0.3\,Z_{\odot}$, the
specific angular momentum of the progenitor allows the loss of the hydrogen
envelope while preserving the helium
core \citep*{woo06b,Fryer_Woosley_Hartmann_1999}. The loss of the envelope reduces
the material that the jet needs to cross in order to escape, while the helium core
should be massive enough to collapse and power a GRB.\\ 

In order to count the number of GRB events in each snapshot, we select from our catalog objects with redshift between $0<z<9.2$, using a procedure similar to that described in section 3 of Campisi et al. 2009, we count all the possible GRB events in the simulated catalog in 3 different subsamples:\\
-{\textit {GRB1}}, obtained by selecting stars with
 age $< t_c = 5\times 10^7 {\rm yr}$ and $M>20\,M_{\odot}$; \\
-{\textit {GRB2}}, including stars of age $< t_c$, $M>20\,M_{\odot}$ and metallicity $Z\leq\,0.3Z_{\odot}$; \\
-{\textit {GRB3}}, defined by selecting stars with age $< t_c$, $M>20\,M_{\odot}$ and
 metallicity $Z\leq0.1\,Z_{\odot}$.   \\
We compute the number of stars ending their lives as LGRBs, assuming a Salpeter Initial Mass Function (IMF) and that the rate of GRB per SNe is on average (over all cosmic times) of about 1 GRB event every 1000 SNe \citep{Porciani_Madau_2001,lan06}\footnote{We test however in Appendix A a different rate (1 GRB every 10.000 SNe), in order to check how the results change with different assumption.}.

\subsection{Star Formation History}
\label{sec:sfrRate}
The collapsar model links LGRBs to the evolution of massive stars whose
lifetimes are negligible on cosmological scales. If no other condition is
required for producing a LGRB event, then the rate of LGRBs should be an
unbiased tracer of the global star formation in the Universe \citep[e.g.][and
references therein]{Totani_1997,Wijers_etal_1998,mao98,Porciani_Madau_2001,bro02,fyn06,pri06,Savaglio_2006,tot06,pro07,li08a}.
However, both observations and theoretical studies indicate that the
metallicity of the progenitor star plays an important role in setting the
necessary conditions for a LGRB explosion. 
In this case, the rate of LGRBs is
expected to be a biased tracer of the cosmic star formation rate.
\begin{figure}
  \includegraphics[scale=0.45, angle=-90]{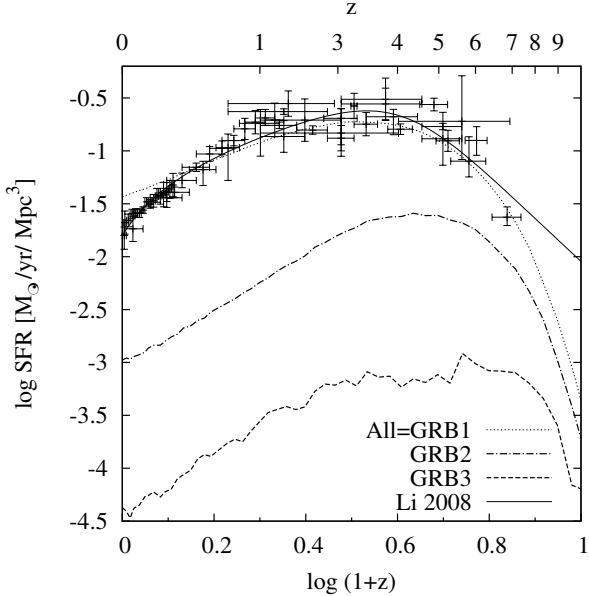}
    \caption{Star Formation Rate History: ${\rm Log}\,SFR\,[M_{\odot}\,{\rm yr}^{-1}\,{\rm Mpc}^{-3}]$ as a function of redshift. Dotted line are the results from GRB1 which also identify all galaxies in the simulation, dot-dashed line is GRB2 and GRB3 is the dashed line. Symbols with error bars are a compilation of observational data (Hawkins 2006), the solid line is a best fit of observational data taken from Li 2008b.}
\label{fig:sfr}
\end{figure}

 In Fig. \ref{fig:sfr} we show the SFH for the simulated and for observed samples .
We compare the cosmic star
formation rate obtained using all galaxies in the simulation box to that
obtained for the three samples defined in the Sec.~\ref{sec:sam} and to the observed SFR in the Universe.
Since normally the SFR calibrations used for deriving the SFH estimates \citep{hop06} are defined {by} assuming the Salpeter (1955) Initial Mass Function, to ensure consistent assumptions throughout, we convert SFH estimates in our simulation\footnote{The simulation adopts the Chabrier IMF} to the Salpeter IMF using a simple scale factor. This scale factor is established by using the Starburst99 code, which models a population-synthesis.

In Fig.1, the sample with no threshold on metallicity (GRB1) traces the global
star formation rate (dotted line in Fig.~\ref{fig:sfr}), this sample also identify all galaxies in the simulation. This is not the case
for the two samples with metallicity thresholds (GRB2 - dot-dashed line and
GRB3 - dashed line). In the GRB2 and GRB3 samples, the LGRB rate peaks at
higher redshift than the cosmic star formation rate, as a consequence of the
global decrease of metallicity with increasing redshift. 
 Data and the best fit (solid line) are taken from Hapkins (2006) and Li (2008b). The SFH of simulation is in agreement with data within their errors bars. As described in Campisi et al. 2009, the peaks of the SFH for the GRBs subsamples are shifted at higher redshift, due to the selection methods since the objects at higher redshift have lower metallicity. \\ 

The important goal of Fig. \ref{fig:sfr} 
is to shed some light on the following issue:
do LGRBs trace the SFR? 
Our results lead to the conclusion that LGRBs are not good tracers, albeit they might be biased tracers of SFR.\\
 But 
all previous works on the study of redshift distribution {of LGRBs} (see sect.1) adopted the assumption, i.e. that the GRB rate is proportional to the SFR, sometimes convolved with a function constraining a metallicity cut-off (e.g \cite{sal07,dai06}).
We show that selecting in our simulation burst with different progenitor's metallicity, the GRB rate follows a different evolution with respect to the SFH. Here, the results of the work which used a strong correlation between the SFR and the LGRB rate could be wrong and should be used with caution. 

The results shown in
Fig. 1 are in qualitative agreement with recent observational
estimates \citep{Kistler_etal_2008}, and with recent theoretical studies also
based on the collapsar model (\citealt*{Yoon_Langer_Norman_2006} and others).

\section{Observed distribution of LGRBs}
To predict the observed distribution of the redshift of LGRBs, we should take into account that only brightest and pointing toward us burst will be observed, so we need to include two important effects: \\
I) the collimation and beaming effects;\\
II) the fraction of GRBs seen by the detector (or luminosity function of GRBs).\\
The number of observed LGRBs is given by:
\begin{equation}
\label{n1}
N_{\rm obs} \sim N_{\rm real} \,\, f_b \,\, \int^\infty _{L_{F_{\textrm{lim}}}} \Phi(L) dL ;
\end{equation}
where $N_{\rm real}$ is the total rate of LGRBs in the simulation, $f_b$ is the fraction of LGRBs pointing toward us and the integral {gives} the fraction of LGRBs with luminosity bigger than the corresponding limit flux of the detector.

\subsection{Beaming effect}
There is a general consensus that GRBs are jetted sources \citep{wax98, rho97}. This implies
fundamental corrections
to the energy budget and the GRB rates. 
A canonical GRB does not light up the full celestial sphere but rather a fraction,
the so called beaming fraction \citep{sar98}:
$$f_b = (1-\cos\,\theta) \sim \theta^2/2 $$
where $\theta$ is the opening angle of the jet. 
Thus, the overall GRB rate clearly depends
on the fact that GRBs are beamed and the rates have to
be corrected by a factor $f_b$. 
Thus the true rate integrated over a time interval is $N_{\rm real}=(f_b)^{-1} N_{\rm obs}$.
 This has been computed traditionally in terms of the beaming correction
factor, which is defined as the ratio of the total number of bursts to the observed ones.
To estimate the overall GRB rate we use the average beaming correction $f_b$.
The average value of $\theta$ 
is $\sim 6\, deg$ \citep{ghi07}, giving $<f_b> \sim 0.0055$. We will use this 
average beaming factor throughout our work.

\subsection{LGRB luminosity function}
The luminosity function (LF) of LGRBs is still poorly tested
as the data are too sparse for an empirical determination of the burst luminosity function. The standard approach to constrain the GRB LF from observations (Porciani $\&$ Madau 2001, Daigne et al. 2006, Salvaterra et al. 2007, and others) is first to assume a model for the LF, for the GRB rate, and for the energy spectrum. Secondly the model parameters are constrained by the observed data. To this aim it is customary to fit the GRBs observed by BATSE
using the differential peak flux distribution, $logN-logP$ diagram (Schmidt 1999).
In particular we fit the observed rate of burst with observed peak fluxes F
 between ($F_1,F_2$), described by the equation:
\begin{eqnarray}
\label{number}
\frac{dN}{dt}(F_1<F<F_2)&=&\int_0^{\infty} dz \frac{dV(z)}{dz} 
\frac{\Delta \Omega}{4\pi} \frac{R_{\rm GRB}(z)}{1+z} \nonumber \\
& \times & \int^{L(F_2,z)}_{L(F_1,z)} dL^\prime \phi(L^\prime)\epsilon(F),
\end{eqnarray}
where ($dV(z)/dz$) is the comoving volume element\footnote{ $dV(z)/dz=4\pi c d_L^2(z)/[H(z)(1+z)^2]$}, $\Delta \Omega$ is the solid angle covered on the sky by the survey, the factor $(1+z)^{-1}$ accounts the cosmological time dilation, $R_{\rm GRB}(z)$ is the 
comoving GRB rate density, $\phi(L^\prime)$ is the GRB luminosity function, and $\epsilon(F)$ is the detector efficiency as a function 
of photon flux.

We fit equation (\ref{number}) using the rate of GRB described in section \ref{sec:sam}, and we assume the following models for the LF and for the energy spectrum. 

\subsubsection{Luminosity function}
To model the number of GRBs at different flux limits, we assume that the luminosity function has the form:
\begin{equation}
\label{Lfun}
\Phi(L) =\,K\,
\left ( \dfrac{L}{L_*} \right ) ^{\xi} \exp \left (-\dfrac{L_*}{L} \right )
\end{equation}
where L is the isotropic equivalent intrinsic burst luminosity, $\xi$ is the asymptotic slope at the faint end, $L_*$ is the characteristic cutoff luminosity, and $K$ is the normalization constant so that the integral over the luminosity function equals unity. We take $K=[L_0\Gamma(-\xi-a)]^{-1}$ (for $\xi<$-1) (Porciani $\&$ Madau 2001).
For the cutoff luminosity we consider differents scenarios, since $L_*$ could to increase with redshift follow the equation $L_*=L(z)=L_0 (1+z)^\delta$.

\subsubsection{Energy spectrum}

We assume the empirical form for the GRB spectrum proposed by Band et al. 1993:
\begin{gather}
\scriptsize
\label{spettro}
 S(E)\propto \\
\left\{ \begin{array}{ll}
\displaystyle{
\left(\frac{E}{100\,{\rm keV}}\right)^\alpha\exp\left[\frac{E
(\beta-\alpha)}{E_b}\right]}
& \textrm{$E<E_b$}, \\
\displaystyle{
\left(\frac{E_b}{100\,{\rm keV}}\right)^{\alpha-\beta} \nonumber 
\exp{(\beta-\alpha)} 
\left(\frac{E}{100 \,{\rm keV}}\right)^\beta}
& \textrm{$E\geq E_b$}.\;
\end{array} \right.
\end{gather}

We adopt the best fit energy spectral indices (i.e., $\alpha \sim -1$ and $\beta \sim -2.25$) 
reported in Preece et al. (2000), and the spectral break energy distribution at $E_{\textrm{b}}=511$ keV (Porciani $\&$ Madau 2001). 
{These parameters were found by fitting 5500 different spectra, which is the most extensive GRB sample with spectral characteristics to date.}

The photon flux $F$[ph/cm$^2$/s] observed at the Earth in the energy band $E_{\textrm{min}}<E<E_{\textrm{max}}$, emitted by an isotropically radiating source at redshift z can be written by:\\
\begin{equation}
\label{peak}
F=  \dfrac{(1+z) \int^{(1+z)E_{\textrm{max}}}_{(1+z)E_{\textrm{min}}}S(E)\,dE }{4 \pi D_L^2},
\end{equation}
where $D_L$ is the luminosity distance and $S(E)$ is the rest frame energy spectrum. We consider $E_{\textrm{min}}$=50 keV and $E_{\textrm{max}}$=300 keV for BATSE, while for the Burst Alert Telescope (BAT, \cite{bar05})
we use $E_{\textrm{min}}$=15 keV and $E_{\textrm{max}}$=150 keV.
It is customary to define the isotropic equivalent burst luminosity in the rest frame photon energy $30-2000$ keV by:\\
\begin{equation}
L=  \int^{2000\,\textrm{keV}}_{30\,\textrm{keV}} E\,S(E)\,dE,
\end{equation}
Thus, combining equations (4-5-6) we get $L(F,z)$ to use in the integration limit of eq. (2).

\subsection{Best fit: results}
In our model we have differents parameters to fit: the characteristic cutoff luminosity $L_*$, the slope $\xi$ of the LF and a third parameter $\delta$ which is the evolution of $L_*$ with redshift, and we fixed it to values between 0-3.5 \citep{sal09}.
We follow the approach of Porciani $\&$ Madau (2001), using the observed differential number counts of BATSE in the range 50-300 keV from Tab. 2 in \cite{kom00}.
The observed sample include 1998 GRBs with peak flux in the range $0.18-20.0\,$ [ph/cm$^2$/s], and detector efficiency described by $\epsilon (F)=0.5[1+{\rm erf}(-4.801+29868\,F)]$ \citep{kom00}. Dai (2009) showed that the choice of the BATSE sample with respect to the \sw\ one is equivalent since the two samples represent the same population of bursts. Moreover, Dai shows that the distribution of the \sw\ sample matches that of the BATSE sample (when approaching the detection limits) so in the follow we can use the same trigger efficiency for \sw\ triggers.\\
The fits to the data are done by minimizing the difference of the $logarithm$ between model and observational data.
This is like a simple
$\chi^2$ minimization, but the points are not independent. We tried to
fit the data directly, finding that the best fit value gave too much weight to the
central regions. We fitted in $logarithm$ values so that the
overall shape of the contours has an increased influence on the
fit.
In all considered cases we always find a clear minimum. 
\begin{figure}
  \centering
  \includegraphics[scale=0.4, angle=-90]{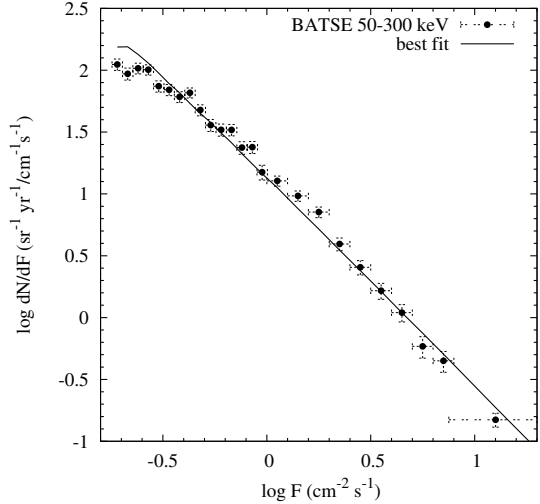}
  \caption{Comparison between BATSE and best fit model for the $log\,N-log\,P$ distribution. The dots are the observed BATSE LGRBs in the 50-300 keV band (Kommers et al. 2000), and the dark line is the predicted distribution with the best fit parameters (we show only for one subsamples, similar fits are obtained for the others cases).}
\label{fitplot}
\end{figure}

The best-fit model is shown in Fig.\ref{fitplot} and the parameters for all the subsamples are listed in Table 1. In Fig.\ref{fitplot} we show only a comparison between the observed distribution of LGRBs and the predicted distribution obtained using the GRB2's subsample, similar lines are obtained with the others subsamples and models. The observed distribution is taken from Kommers et al. (2000), for BATSE detector (energy range 50-300 keV). The data are converted into rates per unit time per unit solid angle following Kommers's work. The horizontal and vertical error bars on data represent the size of the energy bin and Poisson uncertainties, respectively.\\
Table 1 shows the best fit parameters for all subsamples. The errors we quoted are the rms (root mean square) spread in errors from
fitting boostrap catalogue in the same way.\\ 
The given error bars confirm that the characteristic luminosity cutoff remains better determined when there is an evolution of the LF with redshift. We note that for {\emph{all subsamples}} and with very different value of $\delta$, the slope $\xi$ of the LF is well constrained in a range of values between $1.686< - \xi<1.838$ (except one case). These values are in agreement with the results in the literature on the slope of the LF, (e.g. in Daigne et al. 2006, where $1.52<\xi<1.7$). Conversely, the characteristic luminosities span a large range of values $(0.038<L_0<77.875)\times 10^{50}$[erg/s], adopting higher values when we consider the GRB3 subsample, in particular when the LF is constant with the redshift.
 
\begin{table}
\begin{center}
\begin{tabular}{lcc}
\hline
\hline
\multicolumn{3}{c}{Sample GRB1} \\
\hline
$\delta$  & $L_0/(10^{50}{\rm \;erg\;s}^{-1})$ & $-\xi$ \\
\hline
0 & 5.132$\pm$0.791 &1.838$\pm$0.061 \\
1.5   &0.665 $\pm$0.079 &1.726 $\pm$0.039   \\
2.0  &0.347 $\pm$0.039 &1.709 $\pm$0.036   \\
2.5  &0.181 $\pm$0.019 &1.699 $\pm$0.034   \\
3.0  &0.095 $\pm$0.013 &1.694 $\pm$0.069   \\
3.5  &0.049 $\pm$0.015 &1.692 $\pm$0.176   \\

\hline
\hline
\multicolumn{3}{c}{Sample GRB2} \\
\hline
$\delta$ & $L_0/(10^{50}{\rm \;erg\;s}^{-1})$ & $-\xi$  \\
\hline
0 &6.973$\pm$2.066 &1.763$\pm$0.430\\
1.5   & 0.742$\pm$0.088 &1.702$\pm$0.035  \\
2.0  & 0.355$\pm$0.041 &1.694 $\pm$0.033   \\
2.5  & 0.171$\pm$0.019 &1.688 $\pm$0.032   \\
3.0  & 0.081$\pm$0.009 & 1.686$\pm$0.032   \\
3.5  & 0.038$\pm$0.007 & 1.686$\pm$0.066   \\
\hline
\hline
\multicolumn{3}{c}{Sample GRB3} \\
\hline
$\delta$  & $L_0/(10^{50}{\rm \;erg\;s}^{-1})$ & $-\xi$  \\
\hline
0& 77.875$\pm$19.243 & 2.135$\pm$0.184 \\
1.5  & 4.574$\pm$0.410 & 1.820$\pm$0.047  \\
2.0  & 1.970$\pm$0.157 & 1.787$\pm$0.042   \\
2.5  & 0.865$\pm$0.064 & 1.768$\pm$0.039  \\
3.0  & 0.386$\pm$0.028 & 1.760$\pm$0.038 \\
3.5  & 0.173$\pm$0.022 & 1.760$\pm$0.041\\
\hline
\hline
\label{bestfit}
\end{tabular}
\end{center}
\caption{Best--fit parameters for different GRBs' subsamples (see section 2) and different luminosity evolution with $L*=L(z)=L_0 (1+z)^\delta$. 
Errors are computed using boostrap technique.}
\end{table}

\section{LGRB redshift distribution: results}
About 458 GRBs have been detected by the \sw\ satellite since its launch in 2004 November until August 2009.
Among these $\sim$150 have spectroscopic or photometric redshift determination. The number of GRBs with redshift is tightly linked with observing conditions, as explained in \cite{jak06} (hereafter J06). J06 suggested that in order to study the redshift distribution of GRBs we should use a subset of all GRBs well placed for optical observations. This can be achieved by following 6 criteria, necessary to ``clean'' the sample: 1) the burst should have an X-ray position made public within 12 hours; 2) the Galactic foreground should be low, i.e. $A_V<0.5$; 3) the burst should be $>55^\circ$ from the Sun; 4) the burst should be not at a polar declination, $|dec|<70^\circ$; 5) the burst has to be localised with the XRT; and 6) no nearby bright star. Imposing these restrictions does not bias the
sample towards optically bright afterglows; instead each GRB in
the sample has favourable observing conditions, i.e. useful follow-up
observations are likely to be secured.

\begin{table}
\begin{center}
\begin{tabular}{|l||c|c|c|c|c|c|c|}
\hline
\hline
& \multicolumn{5}{c}{$\delta$} \\
  &0.&1.5&2.0&2.5&3.0&3.5   \\
\hline
GRB1&1.71&1.89&1.98&2.08 &2.19&2.33 \\
GRB2&2.43&2.66&2.77&2.90 &3.04&3.19 \\
GRB3&3.12&3.18&3.22&3.27&3.34&3.44 \\
\hline
\multicolumn{6}{c}{\sw\ $<z>\sim\,2.28$} \\
\hline
\hline
\label{bestfit}
\end{tabular}
\end{center}
\caption{Mean redshift for every subsample of GRB with different LF evolution, compared with the mean \sw\ redshift.}
\end{table}

\begin{figure*}
  \centering
  \includegraphics[scale=0.45, angle=-90]{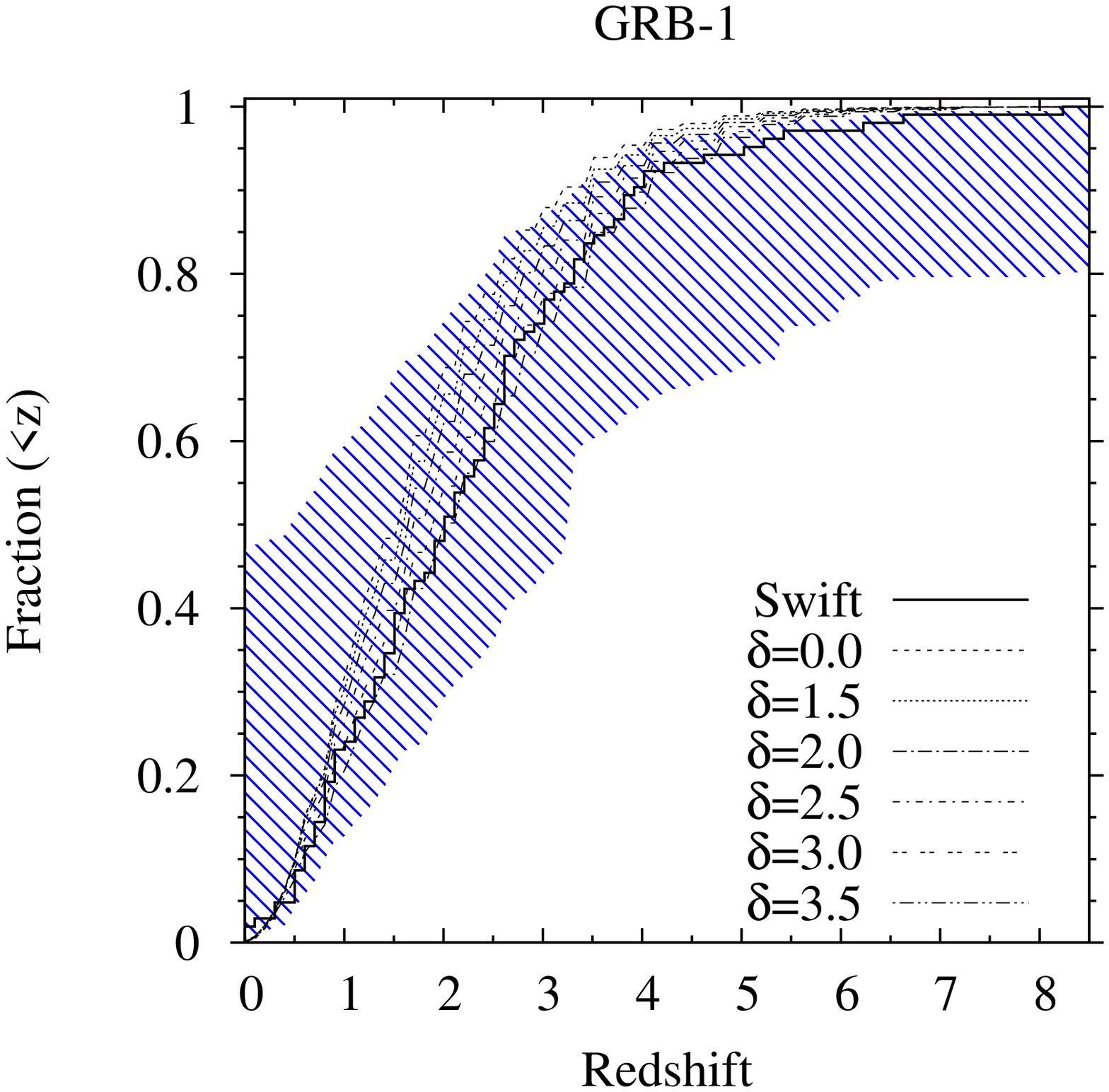}
  \includegraphics[scale=0.45, angle=-90]{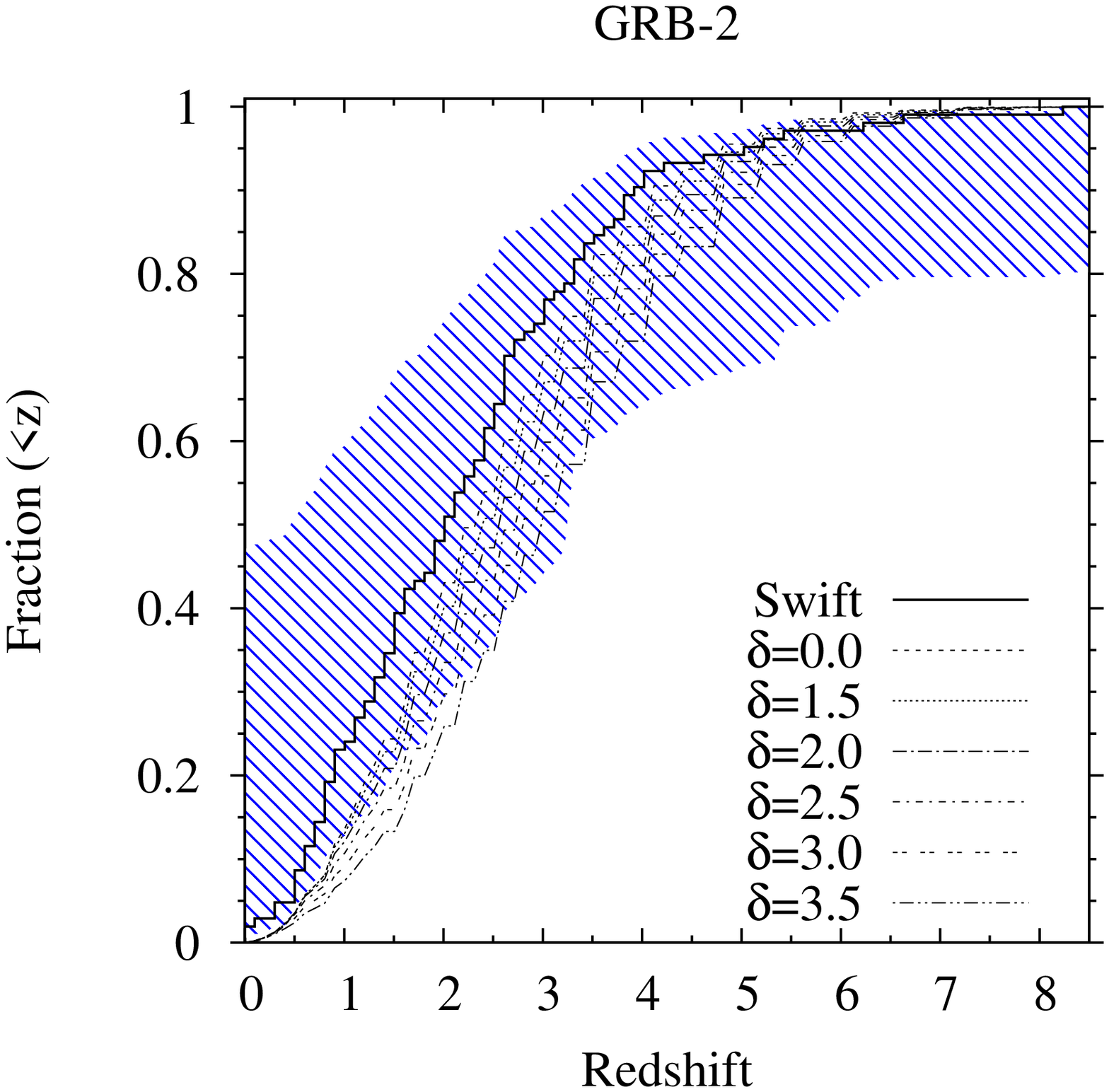}
  \includegraphics[scale=0.45, angle=-90]{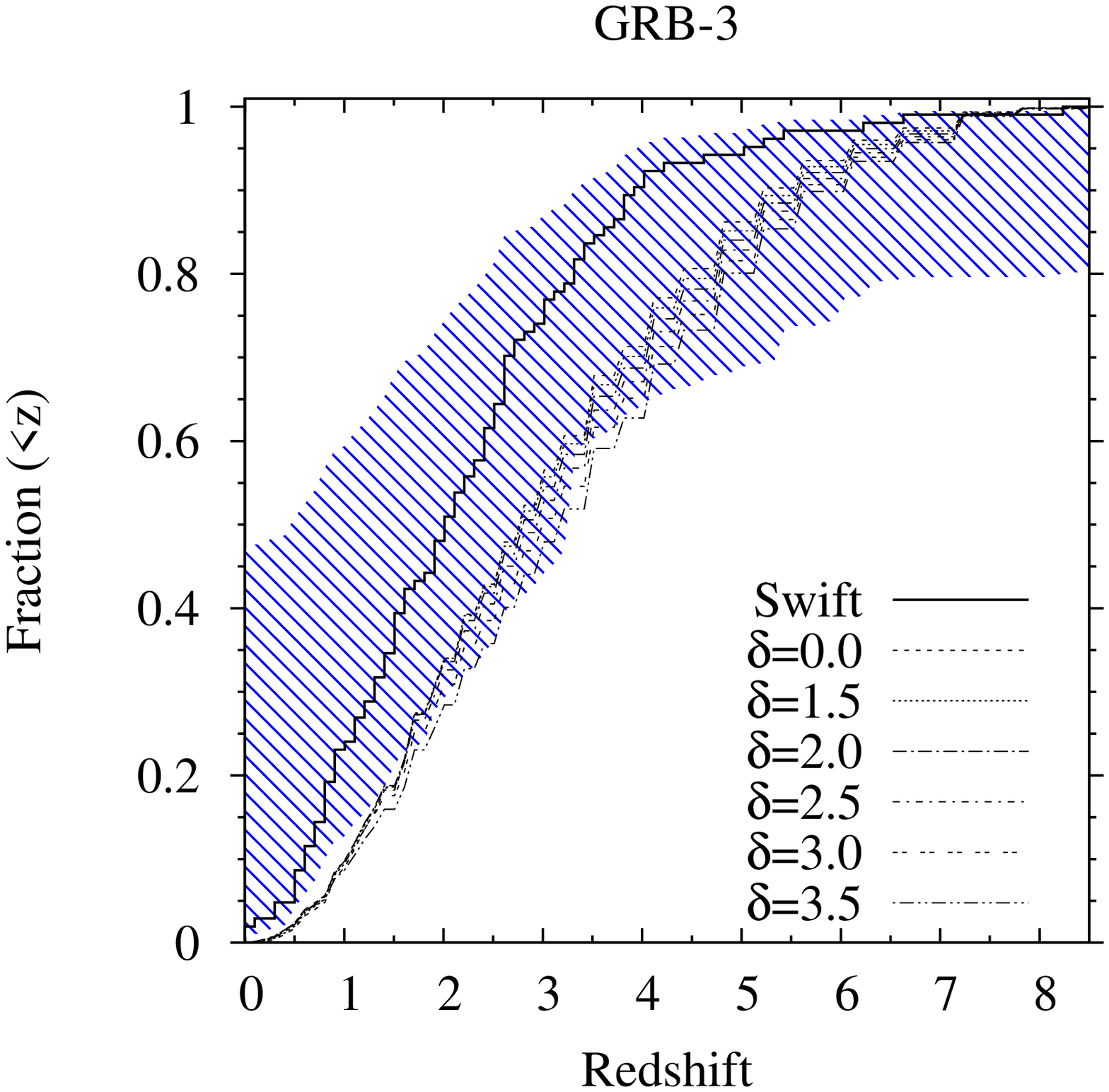}
  \caption{Redshift distribution for LGRB. Thick step is the observed distribution of \sw\ burst with sure measured redshift (J06). The blue area takes in consideration the error region for the steps, following the procedure of Jakobsson et al. 2009. The upper envelope is produced placing GRBs without redshift and those with redshift upper limits at $z=0$. The lower envelope placing the GRBs without firm redshift at the maximum redshift they can have (giving their bluest photometric detection). The model for the expectation of the redshift distribution from our simulation are the dashed lines. For progenitor stars without cut in metallicity (GRB1) and with metallicity lower than $0.3\,Z_{\odot}$ (GRB2) and
$0.1\,Z_{\odot}$ (GRB3). Results are shown for the model with luminosity evolution between 0-3.5.}
\label{fig:rate}
\end{figure*}

Our best-fitting parameters for the LF of GRBs (shown in table 1) are used to predict the redshift distribution for the \sw\ case. We use equation \ref{number} to compute the model prediction of the number of GRB with $L>L(L_{\textrm{lim}},z)$, considering values of F corresponding to $F_{\textrm{lim}}=0.2\,$ [ph/cm$^2$/s] (in order to compare with Daigne et al. 2006).\\
 In Fig. \ref{fig:rate} we compare our model predictions with the number of burst detected by the \sw\ satellite, following the prescriptions of the analysis performed by J06 and by using the updated catalog\footnote{http://www.raunvis.hi.is/$\sim$pja/GRBsample.html} until GRB 090812. We assume that the observed sample of GRBs with redshift determination is representative of all bursts, within the error area \citep{jak09}. Fig. \ref{fig:rate} shows the cumulative redshift distribution of observed and simulated GRBs for the 3 different subsamples.  
In the model, the expected redshift distribution depends on the assumption made on GRB progenitors but also by the evolution of the LF with redshift. The distribution in the case of the GRB1 sample is not so far from the observed one. In particular {it seems} that GRB1 subsample with high evolution for the LF ($\delta>2.5$) reproduce the data.
The observed \sw\ distribution lies also close to the distribution of the GRB2 sample without evolution in $L_*$, more at high redshift. This {implies} that the properties of the GRBs do not change with the redshift, since in our simulation the progenitor's characteristic does not evolve.\\ 
 Figure \ref{fig:rate} shows also that the GRB3 subsample is not a good reproduction of observed data, in particular at low redshift ($z<1.5$) where the lines are outside the error area. 

For completeness in Tab. 2 we also show the comparison between the average value of the redshift for the updated GRBs' sample of J06 and the simulated one. We predict that the value of the $<z>$ evolves with the threshold in metallicity, in agreement with the SFR evolution for the three subsamples. In fact, while the GRB1 prediction gives an average redshift between $1.71-2.33$ (for different values of $\delta$), GRB2 and GRB3 have higher values up to $<z>\sim3.44$ for the extreme case where $\delta=3.5$ and $Z<0.1 Z_{\odot}$.\\
However the observed LGRB have an average redshift of $<z>\sim2.27$, which is more lower than the one predicted from the GRB3 model.
From Fig.3 and Tab.2, we are able to rule out the GRB3 subsample and we conclude that either GRB1 with evolving LF and GRB2 with a non-evolving LF are possible model within error bars.

\subsection{Bright and Faint LGRBs}
Also if there would be an evolution of GRB LF with time (high-z GRB are typically brighter than {low-z} ones)
from Fig.3. seems to need a very strong evolution to reconcile GRB1 with observed data ($\delta>2.5$).
This evolution should imply that the properties of GRBs change more with the 
redshift. 
This scenario {seems} to be unrealistic.

\begin{figure}
  \centering
  \includegraphics[scale=0.4, angle=-90]{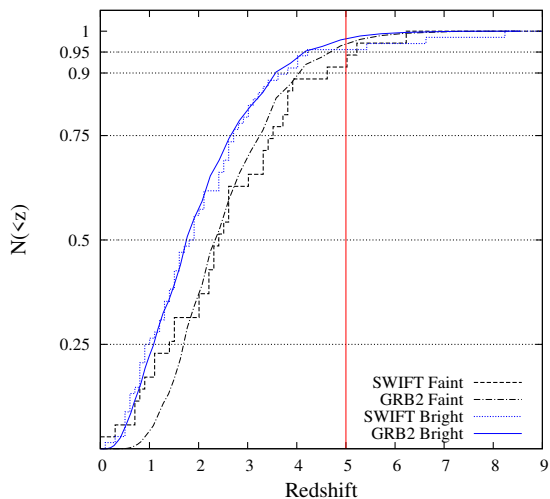}
 \caption{Redshift distribution for \sw\ observed and expected burst for {\sl Faint} burst with peak flux $F \leq1\,$[ph/cm$^2$/s] and {\sl Bright} LGRB with $F>1\,$ [ph/cm$^2$/s]. The lines are the simulations and the steps are the \sw\ data.}
\label{fig:4}
\end{figure}

In order to step over the detection problem, following the same approach of Daigne et al. (2006), we defined two subsamples of \sw\ bursts by selecting those with peak flux $F_{\textrm{lim}}>1.0\,$ [ph/cm$^2$/s] ({\sl Bright}) and those with $F_{\textrm{lim}} \leq 1.0\,$ [ph/cm$^2$/s] ({\sl Faint}). \\
In Fig.\ref{fig:4} we show the cumulative redshift distribution of observed and expected {bursts}. We decided to show only the GRB2 subsample, without evolution in LF, since it seems to be the more realistic model from Fig.3.\\
It is evident that {\sl Bright} and {\sl Faint} objects have different redshift distribution, both for expected and observed {ones}.
We note that the {\sl Bright} observed sample almost overlaps the simulated distribution of bursts, overcoming the problem of Fig.3.
 Conversely, there are more {\sl Faint} bursts at low redshift than predicted in the simulation. 
Both {\sl Bright} and {\sl Faint} have similar behavior a redshift $z>6$. However we argue from Fig.4 that GRB2 subsample with a non-evolving LF is the best possible model to reproduce \sw\ data.

We quantify the probability to find burst a redshift $z>5$ and $z>6$ in Tab. 3 for both subsamples. 
\begin{table}
\begin{center}
\begin{tabular}{|lc|c||c|c|}
\hline
\hline
Rate&\multicolumn{2}{c}{{\sl Faint}}  &\multicolumn{2}{c}{{\sl Bright}}\\
$\%$&$z>5$&$z>6$ &$z>5$&$z>6$\\
\hline
GRB2&$\sim3.8$&$\sim1.1$ &$\sim2.2$& $\sim0.9$\\
Obs& $\sim8.5$&$\sim2.8$ &$\sim4.4$&$\sim2.9$\\
\hline
\hline
\label{fraction}
\end{tabular}
\end{center}
\caption{Fraction of GRBs with redshift $z>5$ for GRB2 subsample without evolution for the LF, and observation data from J06, for burst with peak flux $F\leq 1\,$ [ph/cm$^2$/s] and with $F>1\,$ [ph/cm$^2$/s].}
\end{table}
We expected from model to have about $1\%$ of GRBs with redshift $z>6$. For both cases we find results lower than Daigne et al. (2006), which found that at $z>6$ the fraction of {\sl Bright} {bursts} should be $\sim2-6\%$.
The difference between our estimation and Daigne's one is that their SFR3 (preferreed by redshift distribution of observational data) is probably unrealistic as written in the conclusion of Daigne's paper, suggesting that their results provide strong evidence that the properties of GRBs are redshift dependent. Instead in our case we take into consideration of the GRB luminosity function by considering a model where the characteristic luminosity does not change with redshift, {with the progenitor properties unchanged}.

We claim that if the distribution of the observed sample will not change with the increase of the number of observed LGRB, the LF should not have an evolving cutoff among our modelled values. Moreover the correlation between \sw\ bursts and GRB2 sample implies that LGRB should be produced by a very massive star with metallicity $Z<0.3Z_{\odot}$.

\section{Discussion and conclusions}

In this work, we have studied the luminosity function (LF), the comoving rate and the detection rate of Long Gamma-Ray Burst (LGRBs) in the context of a hierarchical model of galaxy formation. 

Assuming the collapsar model and
imposing different metallicity constraints
 we find that:\\
  \begin{itemize}
\item GRBs with low metallicity progenitors ($Z<0.1-0.3Z_{\odot}$) do not represent a
perfect tracer of the cosmic star formation history (see Fig.1). The deviation of the LGRB rate from the star formation
rate decreases with increasing redshift (as a consequence of the
global decrease of metallicity with increasing redshift) and the bias is stronger as the
metallicity threshold assumed is lowered.\\

\item The LF of LGRBs is well descripted by a power-law with exponential cut-off, with well determined slope between $1.686<\xi<1.838$. Conversely, the characteristic luminosity spans a large range of values ($0.038<L_0<77.875\times 10^{50}$[erg/s]), increasing at lower metallicity threshold (see table 1).\\

\item It is possible to reproduce, within error bars, the redshift distribution of a subset\footnote{Only GRBs well placed for optical observations, \cite{jak06}} of \sw\ data with $F>0.2$ [ph/cm$^2$/s], using: (I) a model without cut in metallicity with a very strong evolving LF, or (II) a model with metallicity threshold $Z<0.3Z_{\odot}$ and a non-evolving LF (see Fig.3). Selecting only {Bright} LGRBs ($F>1\,$[ph/cm$^2$/s]) and 
 suspecting that a scenario where the properties of GRB change so strong with redshift 
seem to be enough unrealistic, we rule out the (I) model (see Fig. 4).\\

\item We predict to have $\sim1\%$ Bright bursts {at} high redshift ($z>6$).\\

\end{itemize}

Our work constrains the LF function of LGRBs, using a rate of GRB not proportional to SFR, and assuming different metallicity for the progenitors, giving us the possibility to assert that the most probable model for LGRB's progenitor have $Z<0.3Z_{\odot}$ and {no} evolution in the LF.

\section*{Acknowledgements}
We are indebted to Dr. Jie Wang for making available their simulated galaxy
catalogues and simulation outputs. MAC thank: Davide Burlon for his main contribution; Shude Mao, Gabriella De Lucia and Ruben Salvaterra for useful discussions; Robert Chapman for helpful comments.
LXL was
supported by the
National Basic Research Program of China (973 Program) under grant
No.2009CB24901. PJ acknowledges support by a Marie Curie European Re-integration Grant
within the 7th European Community Framework Program under contract
number PERG03-GA-2008-226653, and a Grant of Excellence from the
Icelandic Research Fund.

\label{lastpage}
\bibliographystyle{mn2e}
\bibliography{paper}

\begin{appendix}
\section{GRB rate}
\begin{figure*}
  \centering
  \includegraphics[scale=0.45, angle=-90]{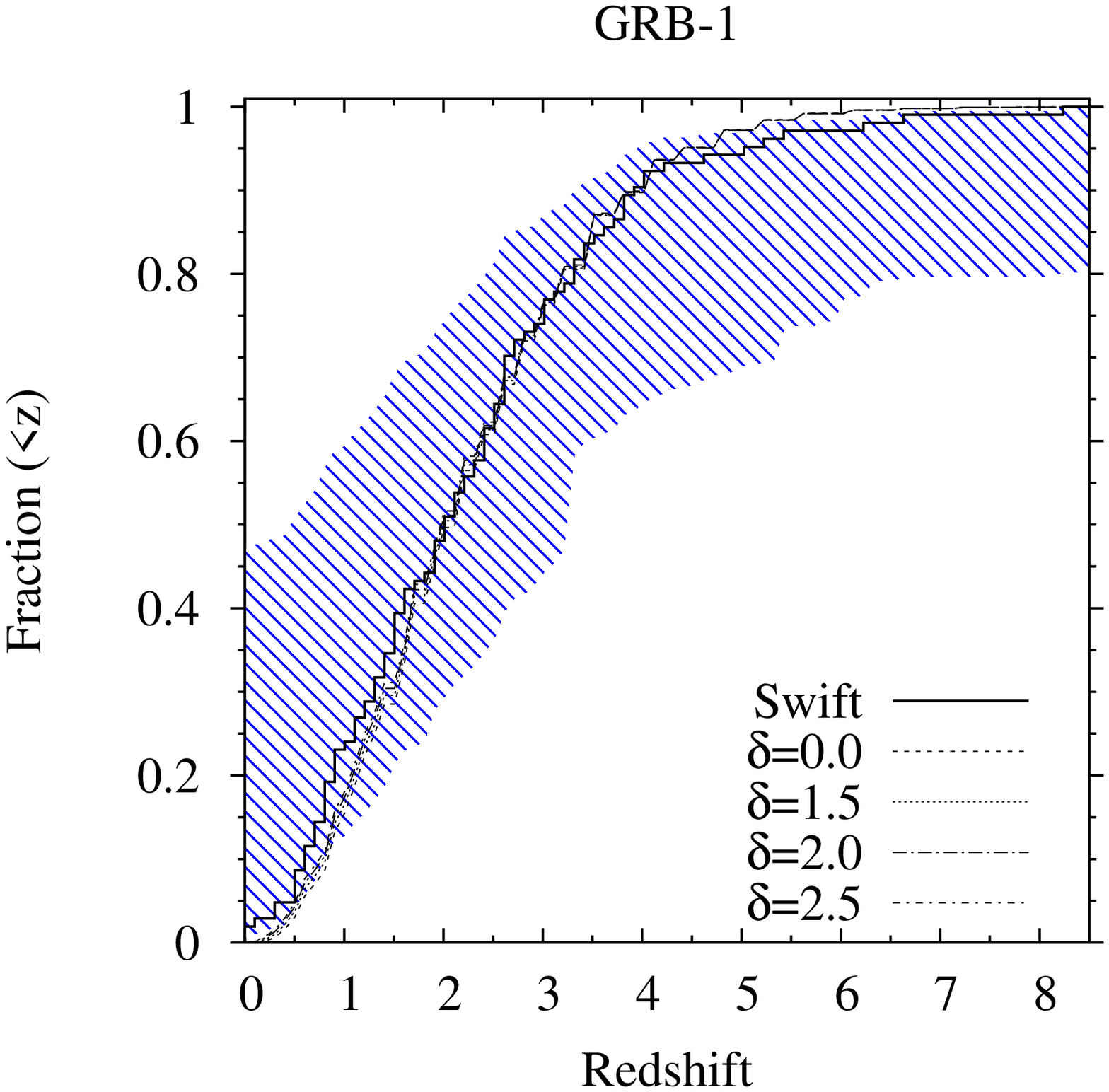}
  \includegraphics[scale=0.45, angle=-90]{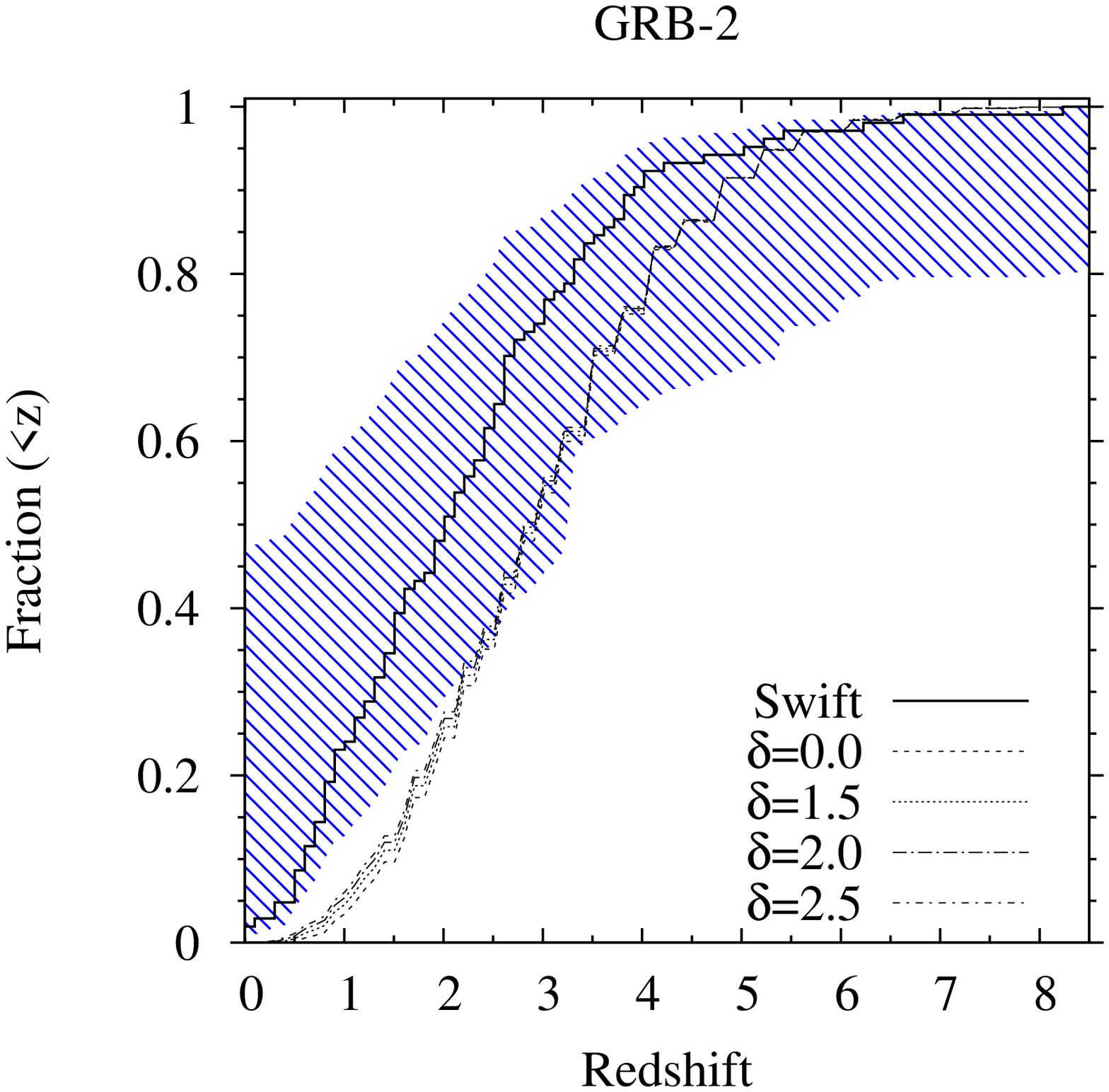}
  \caption{Redshift distribution for LGRB. Thick step is the observed distribution of \sw\ burst with sure measured redshift (J06). The blue area takes in consideration the error region for the steps, following the procedure of Jakobsson et al. 2009. The upper envelope is produced placing GRBs without redshift and those with redshift upper limits at z=0, instead the lower envelope placing the GRBs without firm redshift at the maximum redshift they can have (giving their bluest photometric detection). The model for the expectation of the redshift distribution from our simulation are the dashed lines. For progenitor stars without cut in metallicity (GRB1) and with metallicity lower than $0.3\,Z_{\odot}$ (GRB2). Results are shown for the model with luminosity evolution between 0-2.5.}
\end{figure*}
In order to compute the rate of GRB in our simulation, in section 3 we assume to have 1 GRB every 1000 SNe globally in the Universe, in agreement with \cite{Porciani_Madau_2001,lan06}.
Nevertheless we explore also how change the results assuming a different rate for GRBs. This is what we want test in this appendix.
We repeat all the work, using a rate of 1 GRB every 10000 SNe.
We fit the BATSE sample, obtaining of course differents values for the luminosity function, and we will use this value to compute the redshift distribution. To compare with Fig.3, we show in Fig. A1 the redshift distribution for GRB1 and GRB2 subsample, for GRB3 subsample the rate of GRB is too low to find a best fit for the LF of the BATSE data.\\
The dependence of the LF with the redshift is not remarkable, since (as in GRB3 in Fig.3) the rate of GRB is smaller than previous case. In Fig. A1 {it} is evident that the best model to reproduce the data is the GRB1 subsample (with and without LF evolution). We expect this results because the assumption to have 1 GRB every 10000 SNe calls for a big number of SNe where there are also more low metallicity objects, and for this reason the redshift distribution of GRB1 and GRB2 is shifted at high redshift respect Fig.3.\\
We conclude that, since models and observations
suggest that a metallicity dependency is required, a rate smaller than that using in our work could be unrealistic.
\end{appendix}

\end{document}